\PassOptionsToPackage{table}{xcolor}
\documentclass[]{fairmeta}

\usepackage{amsmath,amssymb}
\usepackage{makecell}

\let\setpaperabstract\abstract
\long\def\abstract#1\end#2{\endgroup\setpaperabstract{#1}}

\title{Tokenizing Numerical and Embedding Features\\ for LLM RecSys}

\author[1]{Zhe Xu}
\author[1]{Ankit Peshin}
\author[1]{Chiyu Zhang}
\author[1]{Feng Qi}
\author[1]{Johnson Lui}
\author[1]{Anil Ramakrishna}
\author[1]{Justin Johnson}
\author[1]{Carl Hu}
\author[1]{Kaushik Rangadurai}
\author[1]{Luke Simon}

\affiliation[1]{Meta}
\correspondence{\email{zhexu@meta.com}}

\begin{abstract}
Large language models (LLMs) are increasingly used as backbone architectures for recommender systems because of their strong sequence modeling and representation learning capabilities. However, most LLM-based recommenders operate primarily on discrete textual tokens, whereas practical recommendation pipelines also rely on continuous numerical features and dense embedding features produced by upstream feature engineering or pretrained encoders. This mismatch limits the ability of LLM-based models to exploit fine-grained non-textual signals. We propose a soft-token fusion framework that maps numerical and embedding features into the LLM embedding space, allowing heterogeneous recommendation signals to be consumed through the standard token interface. We instantiate the framework in a shared-parameter LLM-based two-tower retrieval model and introduce an interaction-based fusion module that refines embedding and numerical soft tokens before they are inserted into the final LLM input. Experiments on three Amazon recommendation benchmarks show that soft-token fusion improves retrieval performance over LLM-based baselines, and that interaction-based fusion is more effective than direct concatenation of heterogeneous soft tokens.
\end{abstract}

\begin{document}

\maketitle

\section{Introduction}
\label{sec:intro}

Large language models (LLMs) have recently emerged as promising backbone models for recommender systems because of their sequence modeling, semantic understanding, and instruction-following capabilities. Recent studies reformulate recommendation as language processing \citep{geng2022p5}, align LLMs with recommendation data through instruction tuning \citep{bao2023tallrec}, and represent items with semantic identifiers for generative retrieval \citep{rajput2023tiger}. These developments indicate a broader shift from task-specific recommendation architectures toward LLM-centered models that consume token sequences and produce textual responses, item identifiers, or ranking signals.

Despite this progress, a key mismatch remains between the token-based input interface of LLMs and the heterogeneous feature representations used in modern recommender systems. Industrial recommendation pipelines commonly rely on categorical IDs, continuous numerical features, and dense embedding features, which are explicitly modeled by neural recommenders through embedding tables, multilayer perceptrons, feature interaction layers, and two-tower retrieval architectures \citep{cheng2016wide,guo2017deepfm,wang2017dcn,naumov2019dlrm,covington2016youtube,yi2019sampling}. Existing LLM-based recommenders often verbalize such signals as text, discretize item representations, or incorporate external features only after LLM encoding. These strategies can obscure scale-sensitive numerical information, discard fine-grained continuous signals, or limit the interaction between textual and non-textual features inside the LLM backbone.

This paper studies how to incorporate numerical and embedding features into LLM-based recommender systems through tokenization in the continuous embedding space. Our key idea is to represent non-textual recommendation features as \emph{soft tokens}. Inspired by continuous prompt learning, prefix tuning, and multimodal LLM adapters \citep{li2021prefix,lester2021prompt,li2023blip2}, we map dense embedding features and continuous numerical features into soft token sequences that can be consumed by the backbone LLM together with ordinary textual tokens.

We instantiate this idea in an LLM-based two-tower retrieval model. The user and item towers encode textual prompts, numerical features, and embedding features into dense retrieval representations. To avoid treating heterogeneous signals as independent side information, we further introduce an interaction-based fusion module that models relations between embedding-derived and numerical soft tokens before they are inserted into the LLM input. This design preserves the standard token-based LLM interface while enabling feature-level interaction within the retrieval backbone.

Our contributions are summarized as follows:
\begin{itemize}
\item We identify the mismatch between token-based LLM inputs and the continuous numerical and dense embedding features commonly used in recommender-system pipelines.
\item We propose a soft-token fusion framework that converts numerical and embedding features into LLM-compatible token representations.
\item We instantiate the framework in a shared-parameter LLM-based two-tower retrieval architecture with interaction-based fusion between heterogeneous soft tokens.
\item We evaluate the method on Amazon recommendation benchmarks and show that interaction-based soft-token fusion improves retrieval performance over LLM-based baselines.
\end{itemize}
\section{Related Work}
\label{sec:related-work}

\subsection{LLM-based Recommender Systems}

LLMs have been increasingly explored as backbone models for recommender systems, mainly due to their semantic understanding, sequence modeling, and instruction-following abilities. Existing studies can be broadly grouped into three directions. The first reformulates recommendation as language modeling or instruction following, where user behaviors, item metadata, reviews, and task descriptions are verbalized into prompts or text-to-text targets \citep{geng2022p5,bao2023tallrec,wu2023survey,xu2024tapping}. The second uses LLMs to enhance recommendation representations, such as generating semantic user or item profiles, augmenting sparse interaction graphs, or improving collaborative representations with textual knowledge \citep{ren2023rlmrec,wei2024llmrec,lyu2023llmrec}. The third studies generative recommendation, where items are represented as textual labels, semantic identifiers, or discrete codes and are generated autoregressively by language models \citep{li2023generative,rajput2023tiger,liao2024llara}. These methods demonstrate the potential of LLMs for recommendation, but most still rely on textualized prompts, item IDs, or discrete semantic codes. In contrast, our work focuses on incorporating continuous-valued numerical features and dense embedding features into LLM-based recommenders through soft-token fusion.

\subsection{Applying Numerical and Embedding Features}

Numerical and embedding features are fundamental inputs in modern recommender systems. Classical and industrial recommendation models explicitly combine sparse categorical embeddings, dense numerical features, and learned user or item representations through feature crossing, attention, and deep interaction modules \citep{cheng2016wide,guo2017deepfm,wang2017dcn,lian2018xdeepfm,naumov2019dlrm,zhou2018din,zhou2019dien,huang2019fibinet,song2019autoint}. Two-tower retrieval models further encode user-side and item-side features into a shared embedding space for efficient large-scale candidate generation \citep{covington2016youtube,yi2019sampling}. Beyond recommendation, tabular Transformer models study how to tokenize structured numerical and categorical features for neural prediction \citep{huang2020tabtransformer,gorishniy2021rtdl}. Soft prompt learning and multimodal LLMs also show that continuous vectors can act as virtual tokens or modality-bridging tokens in pretrained language models \citep{li2021prefix,lester2021prompt,alayrac2022flamingo,li2023blip2}. Our method is inspired by these lines of work but targets a different setting: converting recommendation-specific numerical features and dense embedding features into LLM-compatible soft tokens that can interact with textual tokens inside an LLM-based retrieval backbone.
\section{Preliminaries}
\label{sec:prelim}

\subsection{LLM-based Two-Tower Retrieval Model}

Two-tower retrieval is widely used for large-scale candidate generation in recommender systems \citep{covington2016youtube,yi2019sampling}. Given a user $u$ and an item $i$, the user and item towers map them into a shared embedding space,
\begin{equation}
\mathbf{z}_u=f_{\theta_u}(u), \qquad
\mathbf{z}_i=g_{\theta_i}(i),
\end{equation}
where $\mathbf{z}_u,\mathbf{z}_i\in\mathbb{R}^{d}$. Relevance is measured by an inner product,
\begin{equation}
s(u,i)=\mathbf{z}_u^\top\mathbf{z}_i,
\end{equation}
or by an equivalent similarity function. Since item embeddings can be precomputed and indexed, the model supports efficient approximate nearest-neighbor retrieval at serving time.

In an LLM-based two-tower model, both towers are implemented with language-model backbones. Let $\mathbf{x}_u$ and $\mathbf{x}_i$ denote the tokenized user and item inputs. In our setup, the user and item towers share the same LLM parameters, i.e., $\theta_u=\theta_i=\theta$. We manually append an end-of-sequence token \texttt{[EOS]} to each input sequence and use last-token pooling over the corresponding hidden state. The towers produce hidden sequences and retrieval embeddings by
\begin{equation}
\label{eq:prelim-llm-tower}
\mathbf{H}_a
=
\mathrm{LLM}_{\theta}([\mathbf{x}_a;\mathrm{[EOS]}]),
\qquad
\mathbf{z}_a
=
\mathbf{H}_a[-1],
\qquad a\in\{u,i\},
\end{equation}
where $\mathbf{H}_a[-1]$ denotes the final hidden state at the appended \texttt{[EOS]} token, following standard last-index notation. The input $\mathbf{x}_u$ may contain interaction history, profile features, or behavioral descriptions, while $\mathbf{x}_i$ may contain item title, description, metadata, or other attributes.

The model is typically trained with a contrastive objective using in-batch negatives. For a mini-batch $\mathcal{B}=\{(u_b,i_b^+)\}_{b=1}^{B}$, the loss is
\begin{equation}
\mathcal{L}
=
-\sum_{b=1}^{B}
\log
\frac{\exp(s(u_b,i_b^+)/\tau)}
{\sum_{k=1}^{B}\exp(s(u_b,i_k^+)/\tau)},
\end{equation}
where $\tau$ is a temperature parameter. This objective preserves the scalable retrieval interface of classical two-tower recommenders while using LLMs as semantic encoders.

\subsection{Q-Former}

Q-Former, introduced in BLIP-2 \citep{li2023blip2}, can be viewed as a specialized Transformer that uses a small set of learnable query tokens to extract information from an input feature sequence. Unlike a standard Transformer encoder that updates all input tokens, Q-Former maintains $N_q$ query tokens and lets them attend to external features through cross-attention. Let $\mathbf{Q}\in\mathbb{R}^{N_q\times d}$ denote the learnable query matrix, and let $\mathbf{K}_{\mathrm{in}},\mathbf{V}_{\mathrm{in}}\in\mathbb{R}^{L_v\times d}$ denote the key and value matrices obtained from an input feature sequence. A cross-attention layer updates the queries as
\begin{equation}
\mathrm{Attn}(\mathbf{Q},\mathbf{K}_{\mathrm{in}},\mathbf{V}_{\mathrm{in}})
=
\mathrm{softmax}\!\left(\frac{\mathbf{Q}\mathbf{K}_{\mathrm{in}}^{\top}}{\sqrt{d}}\right)\mathbf{V}_{\mathrm{in}}.
\end{equation}
Stacking such attention layers with feed-forward layers gives
\begin{equation}
\label{eq:prelim-qformer}
\mathbf{T}
=
\mathrm{QFormer}(\mathbf{Q},\mathbf{K}_{\mathrm{in}},\mathbf{V}_{\mathrm{in}})
\in\mathbb{R}^{N_q\times d}.
\end{equation}
The output length is controlled by the number of query tokens rather than by the length of the input feature sequence. Thus, Q-Former acts as both an information bottleneck and a feature adapter: it converts variable-length features into a fixed number of feature-aware tokens. If needed, its outputs are projected to the LLM hidden space as $\widetilde{\mathbf{T}}=\mathbf{T}\mathbf{W}_p$. In this paper, we use Q-Former to convert dense embedding features and numerical feature sequences into compact LLM-compatible soft tokens.

\subsection{Soft Tokens and Numerical Encodings}

Soft tokens are continuous vectors inserted into the input embedding sequence of a language model rather than produced by discrete vocabulary lookup. Given ordinary token embeddings $\mathbf{R}\in\mathbb{R}^{L\times d}$ and soft tokens $\mathbf{S}\in\mathbb{R}^{N\times d}$, the language model can process the concatenated embedding sequence $[\mathbf{R};\mathbf{S}]$ with the same self-attention mechanism used for textual tokens. This provides a convenient interface for injecting non-textual features into an LLM while preserving its token-based architecture.

For scalar numerical features, a direct linear projection may be sensitive to scale and may not capture fine-grained value differences. A common alternative is Fourier feature encoding, which maps a scalar $p$ to sinusoidal features at multiple frequencies:
\begin{equation}
\label{eq:prelim-fourier}
\gamma(p)
=
[\sin(2\pi b_1p),\cos(2\pi b_1p),\ldots,
\sin(2\pi b_Rp),\cos(2\pi b_Rp)].
\end{equation}
The resulting vector can then be projected to the LLM hidden dimension and treated as a numerical feature embedding. In this paper, this encoding is used to represent price values before they are compressed into numerical soft tokens.
\section{Method}
\label{sec:method}

\begin{figure}[t]
\centering
\includegraphics[width=\textwidth]{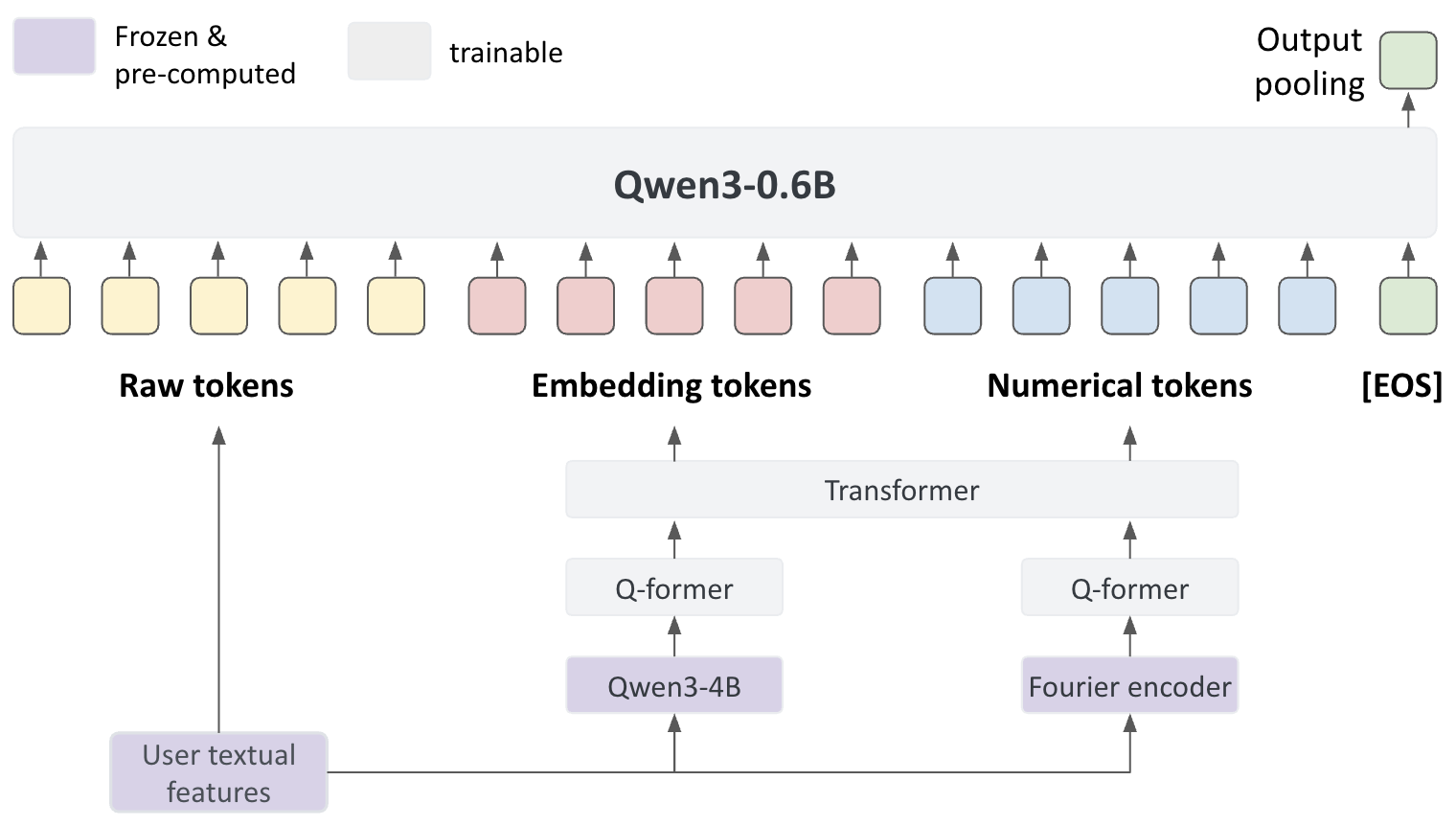}
\caption{Overview of the proposed soft-token fusion framework. Only the user tower is presented for illustration. The same prompt is used in three ways: (1) as text tokens for the final Qwen3-0.6B tower, (2) as dense embedding features encoded by Qwen3-4B, pooled every $G$ tokens, and refined by an embedding Q-Former, and (3) as extracted numerical values that are Fourier-encoded and refined by a numerical Q-Former. The embedding and numerical soft tokens interact before being inserted into the final Qwen3-0.6B tower together with the text tokens and \texttt{[EOS]}.}
\label{fig:method-overview}
\end{figure}

\subsection{Soft Tokens for Embedding Features}

Following the LLM-based tower formulation in Eq.~\eqref{eq:prelim-llm-tower}, let $\mathbf{x}_a=[w_{a,1},\ldots,w_{a,L_a}]$ denote the tokenized prompt for entity $a\in\{u,i\}$. We first encode this prompt with an auxiliary LLM text encoder,
\begin{equation}
\mathbf{H}^{e}_a
=
\mathrm{LLM}_{\mathrm{enc}}(\mathbf{x}_a)
=
[\mathbf{h}^{e}_{a,1},\ldots,\mathbf{h}^{e}_{a,L_a}]
\in\mathbb{R}^{L_a\times d},
\end{equation}
which is precomputed for all users and items. In our implementation, this encoder is Qwen3-4B, while the final retrieval tower uses Qwen3-0.6B.

To reduce storage and computation, we apply mean pooling over every $G$ consecutive token embeddings. With $M_a=\lfloor L_a/G\rfloor$, the $m$-th pooled embedding is
\begin{equation}
\bar{\mathbf{h}}^{e}_{a,m}
=
\frac{1}{G}
\sum_{t=(m-1)G+1}^{mG}
\mathbf{h}^{e}_{a,t},
\qquad m=1,\ldots,M_a,
\end{equation}
yielding $\bar{\mathbf{H}}^{e}_a=[\bar{\mathbf{h}}^{e}_{a,1},\ldots,\bar{\mathbf{h}}^{e}_{a,M_a}]\in\mathbb{R}^{M_a\times d}$. Following the Q-Former adapter in Eq.~\eqref{eq:prelim-qformer}, we set $\mathbf{K}^{e}_a=\mathbf{V}^{e}_a=\bar{\mathbf{H}}^{e}_a$ and compress this sequence into $N_e$ embedding soft tokens:
\begin{equation}
\mathbf{S}^{e}_a
=
\mathrm{QFormer}_{e}(\mathbf{Q}^{e},\mathbf{K}^{e}_a,\mathbf{V}^{e}_a)
\in\mathbb{R}^{N_e\times d},
\end{equation}
where $\mathbf{Q}^{e}\in\mathbb{R}^{N_e\times d}$ is a set of learnable embedding queries.

The Q-Former adapter gives a fixed-length soft-token bottleneck while allowing the learnable queries to attend over variable-length prompt embeddings. Appendix~\ref{sec:appendix-additional-ablation} empirically verifies this design choice by comparing it with an MLP-based alternative.

\subsection{Soft Tokens for Numerical Features}

We also convert numerical features into soft tokens, focusing on price. For a user $u$, the clicked items induce a price sequence $\mathbf{p}_u=[p_{u,1},\ldots,p_{u,T_u}]$; for an item $i$, we define $\mathbf{p}_i=[p_i]$ and $T_i=1$. As introduced in Eq.~\eqref{eq:prelim-fourier}, each scalar price $p$ is encoded with Fourier features \citep{tancik2020fourier,mildenhall2020nerf}:
\begin{equation}
\gamma(p)
=
[\sin(2\pi b_1p),\cos(2\pi b_1p),\ldots,
\sin(2\pi b_Rp),\cos(2\pi b_Rp)],
\end{equation}
where $\{b_r\}_{r=1}^{R}$ are frequency bands. A learnable projection maps this encoding to the LLM hidden dimension,
\begin{equation}
\mathbf{e}^{n}(p)=\phi_n(\gamma(p))\in\mathbb{R}^{d}.
\end{equation}

For $a\in\{u,i\}$, applying this encoder to $\mathbf{p}_a$ gives
\begin{equation}
\mathbf{E}^{n}_a
=
[\mathbf{e}^{n}(p_{a,1}),\ldots,\mathbf{e}^{n}(p_{a,T_a})]
\in\mathbb{R}^{T_a\times d}.
\end{equation}
Again following Eq.~\eqref{eq:prelim-qformer}, we set $\mathbf{K}^{n}_a=\mathbf{V}^{n}_a=\mathbf{E}^{n}_a$ and use a numerical Q-Former to produce $N_n$ numerical soft tokens:
\begin{equation}
\mathbf{S}^{n}_a
=
\mathrm{QFormer}_{n}(\mathbf{Q}^{n},\mathbf{K}^{n}_a,\mathbf{V}^{n}_a)
\in\mathbb{R}^{N_n\times d},
\end{equation}
where $\mathbf{Q}^{n}\in\mathbb{R}^{N_n\times d}$ is a set of learnable numerical queries.

Fourier encoding preserves continuous price variation before the numerical Q-Former compresses the sequence. Appendix~\ref{sec:appendix-additional-ablation} further compares this choice with a tokenizer-based numerical representation.

\subsection{Soft-Token Fusion}

Let $\mathbf{R}_a=\mathrm{Embed}_{\theta}(\mathbf{x}_a)$ denote the prompt-token embeddings under Qwen3-0.6B, and let $\mathbf{e}_{\mathrm{EOS}}$ denote the embedding of the appended \texttt{[EOS]} token. A direct-concatenation baseline inserts both soft-token streams without additional interaction:
\begin{equation}
\mathbf{Y}^{\mathrm{plain}}_a
=
[\mathbf{R}_a;\mathbf{S}^{e}_a;\mathbf{S}^{n}_a;\mathbf{e}_{\mathrm{EOS}}].
\end{equation}
Although LLM self-attention can mix textual, embedding, and numerical tokens \citep{vaswani2017attention}, direct concatenation is not always effective. Empirically, it can underperform single-source variants such as $[\mathbf{R}_a;\mathbf{S}^{e}_a;\mathbf{e}_{\mathrm{EOS}}]$ and $[\mathbf{R}_a;\mathbf{S}^{n}_a;\mathbf{e}_{\mathrm{EOS}}]$.

We therefore introduce a lightweight pre-interaction module for heterogeneous soft tokens. The embedding and numerical soft tokens are concatenated and passed through a Transformer fusion module,
\begin{equation}
[\widetilde{\mathbf{S}}^{e}_a;\widetilde{\mathbf{S}}^{n}_a]
=
\mathrm{Transformer}_{\psi}
([\mathbf{S}^{e}_a;\mathbf{S}^{n}_a],\mathbf{A}),\label{eq:token-fusion}
\end{equation}
where $\widetilde{\mathbf{S}}^{e}_a\in\mathbb{R}^{N_e\times d}$, $\widetilde{\mathbf{S}}^{n}_a\in\mathbb{R}^{N_n\times d}$, and $\mathbf{A}$ is an attention mask. We consider bidirectional, embedding-to-numerical, and numerical-to-embedding masks, corresponding to mutual interaction or one-way cross-feature attention.

The refined soft tokens are then inserted after the text-token embeddings and before the \texttt{[EOS]} embedding,
\begin{equation}
\widetilde{\mathbf{Y}}_a
=
[\mathbf{R}_a;\widetilde{\mathbf{S}}^{e}_a;\widetilde{\mathbf{S}}^{n}_a;\mathbf{e}_{\mathrm{EOS}}],
\end{equation}
and encoded by the shared-parameter Qwen3-0.6B tower to obtain the final retrieval representation, using the same last-token pooling rule as Eq.~\eqref{eq:prelim-llm-tower}:
\begin{equation}
\mathbf{H}_a
=
\mathrm{LLM}_{\theta}(\widetilde{\mathbf{Y}}_a),
\qquad
\mathbf{z}_a
=
\mathbf{H}_a[-1].
\end{equation}
This design keeps the standard token-based LLM interface while explicitly modeling interactions between embedding-derived and numerical soft tokens.
\section{Experiments}
\label{sec:exp}

\subsection{Experimental Setup}

\subsubsection{Datasets}
\label{sec:datasets}

We evaluate our method using the datasets and preprocessing protocol adopted by TIGER \citep{rajput2023recommender}. Specifically, we use three public benchmarks from the Amazon Product Reviews dataset \citep{he2016ups}: \textit{Beauty}, \textit{Sports and Outdoors}, and \textit{Toys and Games}. These datasets contain user reviews and item metadata collected from May 1996 to July 2014, and are commonly used for sequential recommendation evaluation.

Following TIGER, we construct a chronological item sequence for each user by sorting the user's review history by timestamp. Users with fewer than five reviews are removed. We adopt the leave-one-out evaluation protocol: for each user sequence, the last item is used for testing, the second-to-last item is used for validation, and the remaining items are used for training. In our point-wise retrieval setup, the chronological item sequence serves as the user-side input.

The statistics of the processed datasets are summarized in Table~\ref{tab:dataset_statistics}.

\begin{table}[!t]
\centering
\caption{Statistics of the three Amazon review datasets.}
\label{tab:dataset_statistics}
\begin{tabular}{lrrrr}
\toprule
\textbf{Dataset} & \textbf{\# Users} & \textbf{\# Items} & \textbf{Mean Seq. Len.} & \textbf{Median Seq. Len.} \\
\midrule
Beauty & 22,363 & 12,101 & 8.87 & 6 \\
Sports and Outdoors & 35,598 & 18,357 & 8.32 & 6 \\
Toys and Games & 19,412 & 11,924 & 8.63 & 6 \\
\bottomrule
\end{tabular}
\end{table}

\subsubsection{Implementation Details}

\paragraph{Preprocessing.} We convert user and item features into natural-language prompts so that the LLM tower can process recommendation metadata through its standard textual interface. Details of the prompt construction are provided in Appendix~\ref{sec:appendix-preprocessing}.

\paragraph{Backbone Model.}
We use Qwen3-0.6B as the backbone language model for the final two-tower retrieval encoder. Qwen3-0.6B is the smallest dense model in the Qwen3 family, making it suitable for efficient recommendation experiments while still retaining the multilingual and instruction-following capabilities of the Qwen3 series \citep{yang2025qwen3}. We initialize this backbone from the official Qwen3-0.6B checkpoint.\footnote{\url{https://huggingface.co/Qwen/Qwen3-0.6B}}

\subsubsection{Metrics}

We evaluate sequential recommendation performance using standard top-$K$ ranking metrics: \emph{Recall@K} and \emph{Normalized Discounted Cumulative Gain at K (NDCG@K)}, with $K\in\{5,10\}$. Formal definitions are provided in Appendix~\ref{sec:appendix-metrics}.

\subsection{Main Results}

Tables~\ref{tab:beauty_ablation}, \ref{tab:sports_ablation}, and \ref{tab:toys_ablation} present ablation studies on Beauty, Sports and Outdoors, and Toys and Games, respectively. We compare five variants of the LLM-based two-tower model: a text-only baseline, two single-source variants that add either embedding soft tokens or numerical soft tokens, a direct concatenation variant that uses both feature types, and the proposed interaction-based fusion variant. These ablations isolate the contribution of each non-textual feature stream and test whether modeling feature interactions before LLM encoding is more effective than simple concatenation.

\begin{table}[!t]
  \centering
  \caption{Ablation study of LLM-based two-tower variants on the Beauty dataset.}
  \label{tab:beauty_ablation}
  \begin{tabular}{ccccccc}
    \toprule
    \makecell[c]{\textbf{Embedding}\\\textbf{Feature}} & \makecell[c]{\textbf{Numerical}\\\textbf{Feature}} & \makecell[c]{\textbf{Fusion}\\\textbf{Method}} & \makecell[c]{\textbf{R@5}} & \makecell[c]{\textbf{R@10}} & \makecell[c]{\textbf{N@5}} & \makecell[c]{\textbf{N@10}} \\
    \midrule
               &            & N/A                         & 0.0415          & 0.0693          & 0.0239          & 0.0328          \\
    $\checkmark$ &            & N/A                         & 0.0439          & 0.0752          & 0.0266          & 0.0367          \\
               & $\checkmark$ & N/A                         & \textbf{0.0463} & 0.0749          & 0.0275          & 0.0368          \\
    $\checkmark$ & $\checkmark$ & Concatenation               & 0.0423          & 0.0719          & 0.0245          & 0.0340          \\
    \makecell[c]{$\checkmark$} & \makecell[c]{$\checkmark$} & \makecell[c]{Interaction} & \makecell[c]{0.0459} & \makecell[c]{\textbf{0.0766}} & \makecell[c]{\textbf{0.0278}} & \makecell[c]{\textbf{0.0377}} \\
    \bottomrule
  \end{tabular}
\end{table}

\begin{table}[!t]
  \centering
  \caption{Ablation study of LLM-based two-tower variants on the Sports and Outdoors dataset.}
  \label{tab:sports_ablation}
  \begin{tabular}{ccccccc}
    \toprule
    \makecell[c]{\textbf{Embedding}\\\textbf{Feature}} & \makecell[c]{\textbf{Numerical}\\\textbf{Feature}} & \makecell[c]{\textbf{Fusion}\\\textbf{Method}} & \makecell[c]{\textbf{R@5}} & \makecell[c]{\textbf{R@10}} & \makecell[c]{\textbf{N@5}} & \makecell[c]{\textbf{N@10}} \\
    \midrule
               &            & N/A                         & 0.0205          & 0.0358          & 0.0120          & 0.0169          \\
    $\checkmark$ &            & N/A                         & 0.0216          & 0.0365          & 0.0133          & 0.0182          \\
               & $\checkmark$ & N/A                         & 0.0242          & 0.0408          & 0.0146          & 0.0199          \\
    $\checkmark$ & $\checkmark$ & Concatenation               & 0.0220          & 0.0382          & 0.0130          & 0.0182          \\
    \makecell[c]{$\checkmark$} & \makecell[c]{$\checkmark$} & \makecell[c]{Interaction} & \makecell[c]{\textbf{0.0253}} & \makecell[c]{\textbf{0.0410}} & \makecell[c]{\textbf{0.0154}} & \makecell[c]{\textbf{0.0205}} \\
    \bottomrule
  \end{tabular}
\end{table}

\begin{table}[!t]
  \centering
  \caption{Ablation study of LLM-based two-tower variants on the Toys and Games dataset.}
  \label{tab:toys_ablation}
  \begin{tabular}{ccccccc}
    \toprule
    \makecell[c]{\textbf{Embedding}\\\textbf{Feature}} & \makecell[c]{\textbf{Numerical}\\\textbf{Feature}} & \makecell[c]{\textbf{Fusion}\\\textbf{Method}} & \makecell[c]{\textbf{R@5}} & \makecell[c]{\textbf{R@10}} & \makecell[c]{\textbf{N@5}} & \makecell[c]{\textbf{N@10}} \\
    \midrule
               &            & N/A                         & 0.0600          & 0.0945          & 0.0352          & 0.0463          \\
    $\checkmark$ &            & N/A                         & 0.0637          & 0.1008          & 0.0377          & 0.0497          \\
               & $\checkmark$ & N/A                         & 0.0557          & 0.0944          & 0.0327          & 0.0452          \\
    $\checkmark$ & $\checkmark$ & Concatenation               & 0.0575          & 0.0950          & 0.0343          & 0.0464          \\
    \makecell[c]{$\checkmark$} & \makecell[c]{$\checkmark$} & \makecell[c]{Interaction} & \makecell[c]{\textbf{0.0695}} & \makecell[c]{\textbf{0.1064}} & \makecell[c]{\textbf{0.0417}} & \makecell[c]{\textbf{0.0536}} \\
    \bottomrule
  \end{tabular}
\end{table}

The ablation results show that non-textual recommendation features provide useful signals for LLM-based retrieval, but that the fusion strategy is crucial. On Beauty and Sports, numerical soft tokens provide the strongest single-source gains over the text-only baseline, increasing R@5 from 0.0415 to 0.0463 on Beauty and from 0.0205 to 0.0242 on Sports. On Toys, embedding soft tokens are more effective, improving R@5 from 0.0600 to 0.0637 and N@10 from 0.0463 to 0.0497. These trends suggest that the relative utility of embedding and numerical features depends on dataset characteristics.

Directly concatenating embedding and numerical soft tokens is not consistently beneficial: in all three datasets, concatenation underperforms at least one single-source variant, and on Toys it falls below the text-only baseline on R@5. This supports the motivation in Section~\ref{sec:method}: heterogeneous soft tokens can interfere with each other if they are appended to the LLM input without prior interaction. In contrast, the interaction-based fusion variant achieves the best or tied-best performance on most metrics. It obtains the best R@10, N@5, and N@10 on Beauty, and the best scores across all metrics on both Sports and Toys. The gains are especially clear on Toys, where interaction-based fusion improves R@5 from 0.0600 to 0.0695 over the text-only baseline and from 0.0575 to 0.0695 over direct concatenation. Additional design ablations in Appendix~\ref{sec:appendix-additional-ablation} further support the use of Q-Former adapters and Fourier numerical encoding. We therefore use the interaction-based fusion variant as the proposed model in the comparison with published baselines.

\subsection{Comparison with Published Baselines}

To contextualize the proposed model, Tables~\ref{tab:beauty_baseline_comparison}, \ref{tab:sports_baseline_comparison}, and \ref{tab:toys_baseline_comparison} compare the interaction-based variant with published baselines on Beauty, Sports and Outdoors, and Toys and Games, respectively. We report the retained baselines from the TIGER protocol~\citep{rajput2023recommender}, including TIGER itself, using the metrics available in the source paper.

\begin{table}[!t]
\centering
\caption{Comparison with published baselines on the Beauty dataset.}
\label{tab:beauty_baseline_comparison}
\begin{tabular}{lcccc}
\toprule
\textbf{Method} & \textbf{R@5} & \textbf{R@10} & \textbf{N@5} & \textbf{N@10} \\
\midrule
GRU4Rec~\citep{hidasi2016gru4rec} & 0.0164 & 0.0283 & 0.0099 & 0.0137 \\
Caser~\citep{tang2018caser} & 0.0205 & 0.0347 & 0.0131 & 0.0176 \\
HGN~\citep{ma2019hgn} & 0.0325 & 0.0512 & 0.0206 & 0.0266 \\
SASRec~\citep{kang2018sasrec} & 0.0387 & 0.0605 & 0.0249 & 0.0318 \\
BERT4Rec~\citep{sun2019bert4rec} & 0.0203 & 0.0347 & 0.0124 & 0.0170 \\
FDSA~\citep{zhang2019fdsa} & 0.0267 & 0.0407 & 0.0163 & 0.0208 \\
S3-Rec~\citep{zhou2020s3rec} & 0.0387 & 0.0647 & 0.0244 & 0.0327 \\
TIGER~\citep{rajput2023recommender} & 0.0454 & 0.0648 & \textbf{0.0321} & \textbf{0.0384} \\
\rowcolor{gray!10}
Ours (Interaction) & \textbf{0.0459} & \textbf{0.0766} & 0.0278 & 0.0377 \\
\bottomrule
\end{tabular}
\end{table}

\begin{table}[!t]
\centering
\caption{Comparison with published baselines on the Sports and Outdoors dataset.}
\label{tab:sports_baseline_comparison}
\begin{tabular}{lcccc}
\toprule
\textbf{Method} & \textbf{R@5} & \textbf{R@10} & \textbf{N@5} & \textbf{N@10} \\
\midrule
GRU4Rec~\citep{hidasi2016gru4rec} & 0.0129 & 0.0204 & 0.0086 & 0.0110 \\
Caser~\citep{tang2018caser} & 0.0116 & 0.0194 & 0.0072 & 0.0097 \\
HGN~\citep{ma2019hgn} & 0.0189 & 0.0313 & 0.0120 & 0.0159 \\
SASRec~\citep{kang2018sasrec} & 0.0233 & 0.0350 & 0.0154 & 0.0192 \\
BERT4Rec~\citep{sun2019bert4rec} & 0.0115 & 0.0191 & 0.0075 & 0.0099 \\
FDSA~\citep{zhang2019fdsa} & 0.0182 & 0.0288 & 0.0122 & 0.0156 \\
S3-Rec~\citep{zhou2020s3rec} & 0.0251 & 0.0385 & 0.0161 & 0.0204 \\
TIGER~\citep{rajput2023recommender} & \textbf{0.0264} & 0.0400 & \textbf{0.0181} & \textbf{0.0225} \\
\rowcolor{gray!10}
Ours (Interaction) & 0.0253 & \textbf{0.0410} & 0.0154 & 0.0205 \\
\bottomrule
\end{tabular}
\end{table}

\begin{table}[!t]
\centering
\caption{Comparison with published baselines on the Toys and Games dataset.}
\label{tab:toys_baseline_comparison}
\begin{tabular}{lcccc}
\toprule
\textbf{Method} & \textbf{R@5} & \textbf{R@10} & \textbf{N@5} & \textbf{N@10} \\
\midrule
GRU4Rec~\citep{hidasi2016gru4rec} & 0.0097 & 0.0176 & 0.0059 & 0.0084 \\
Caser~\citep{tang2018caser} & 0.0166 & 0.0270 & 0.0107 & 0.0141 \\
HGN~\citep{ma2019hgn} & 0.0321 & 0.0497 & 0.0221 & 0.0277 \\
SASRec~\citep{kang2018sasrec} & 0.0463 & 0.0675 & 0.0306 & 0.0374 \\
BERT4Rec~\citep{sun2019bert4rec} & 0.0116 & 0.0203 & 0.0071 & 0.0099 \\
FDSA~\citep{zhang2019fdsa} & 0.0228 & 0.0381 & 0.0140 & 0.0189 \\
S3-Rec~\citep{zhou2020s3rec} & 0.0443 & 0.0700 & 0.0294 & 0.0376 \\
TIGER~\citep{rajput2023recommender} & 0.0521 & 0.0712 & 0.0371 & 0.0432 \\
\rowcolor{gray!10}
Ours (Interaction) & \textbf{0.0695} & \textbf{0.1064} & \textbf{0.0417} & \textbf{0.0536} \\
\bottomrule
\end{tabular}
\end{table}

Compared with TIGER, the interaction-based two-tower model improves R@10 on all three datasets and achieves the strongest overall performance on Toys and Games. On Beauty and Sports and Outdoors, TIGER remains competitive on ranking-sensitive NDCG metrics, while our model is stronger on R@10. These results show that the proposed LLM-based two-tower retrieval framework is competitive with strong sequential and generative recommendation baselines while using independently encoded user and item representations.
\section{Conclusion}
\label{sec:conclusion}

This paper studied how to incorporate continuous numerical features and dense embedding features into LLM-based recommender systems. Instead of converting these non-textual signals into ordinary text or using them only as post-encoding side information, we represented them as LLM-compatible soft tokens. We instantiated this idea in a shared-parameter two-tower retrieval architecture, where embedding features and numerical price features are compressed by Q-Formers and inserted into the LLM input sequence. To better combine heterogeneous signals, we further introduced an interaction-based fusion module that refines embedding-derived and numerical soft tokens before final LLM encoding.

Experiments on three Amazon recommendation benchmarks show that both numerical and embedding soft tokens can improve LLM-based retrieval, while direct concatenation of heterogeneous soft tokens is not consistently effective. In contrast, interaction-based fusion achieves the strongest overall performance across Beauty, Sports and Outdoors, and Toys and Games, demonstrating the importance of modeling relations among non-textual recommendation features before injecting them into the LLM backbone. Additional design ablations support the use of Q-Former adapters for soft-token construction and Fourier features for continuous numerical encoding. These results suggest that soft-token fusion is a practical interface for bridging LLMs with feature-rich recommender-system pipelines. Future work can extend this framework to additional feature types, larger backbone models, and more realistic large-scale retrieval settings.

\clearpage
\bibliographystyle{assets/plainnat}
\bibliography{ref}

@inproceedings{geng2022p5,
  title = {Recommendation as Language Processing (RLP): A Unified Pretrain, Personalized Prompt \& Predict Paradigm (P5)},
  author = {Geng, Shijie and Liu, Shuchang and Fu, Zuohui and Ge, Yingqiang and Zhang, Yongfeng},
  booktitle = {Proceedings of the 16th ACM Conference on Recommender Systems},
  pages = {299--315},
  year = {2022},
  doi = {10.1145/3523227.3546767}
}

@inproceedings{bao2023tallrec,
  title = {TALLRec: An Effective and Efficient Tuning Framework to Align Large Language Model with Recommendation},
  author = {Bao, Keqin and Zhang, Jizhi and Zhang, Yang and Wang, Wenjie and Feng, Fuli and He, Xiangnan},
  booktitle = {Proceedings of the 17th ACM Conference on Recommender Systems},
  pages = {1007--1014},
  year = {2023},
  doi = {10.1145/3604915.3608857}
}

@inproceedings{cheng2016wide,
  title = {Wide \& Deep Learning for Recommender Systems},
  author = {Cheng, Heng-Tze and Koc, Levent and Harmsen, Jeremiah and Shaked, Tal and Chandra, Tushar and Aradhye, Hrishi and Anderson, Glen and Corrado, Greg and Chai, Wei and Ispir, Mustafa and Anil, Rohan and Haque, Zakaria and Hong, Lichan and Jain, Vihan and Liu, Xiaobing and Shah, Hemal},
  booktitle = {Proceedings of the 1st Workshop on Deep Learning for Recommender Systems},
  pages = {7--10},
  year = {2016},
  doi = {10.1145/2988450.2988454}
}

@inproceedings{guo2017deepfm,
  title = {DeepFM: A Factorization-Machine Based Neural Network for CTR Prediction},
  author = {Guo, Huifeng and Tang, Ruiming and Ye, Yunming and Li, Zhenguo and He, Xiuqiang},
  booktitle = {Proceedings of the 26th International Joint Conference on Artificial Intelligence},
  pages = {1725--1731},
  year = {2017}
}

@inproceedings{wang2017dcn,
  title = {Deep \& Cross Network for Ad Click Predictions},
  author = {Wang, Ruoxi and Fu, Bin and Fu, Gang and Wang, Mingliang},
  booktitle = {Proceedings of the ADKDD'17},
  pages = {12:1--12:7},
  year = {2017},
  doi = {10.1145/3124749.3124754}
}

@article{naumov2019dlrm,
  title = {Deep Learning Recommendation Model for Personalization and Recommendation Systems},
  author = {Naumov, Maxim and Mudigere, Dheevatsa and Shi, Hao-Jun Michael and Huang, Jianyu and Sundaraman, Narayanan and Park, Jongsoo and Wang, Xiaodong and Gupta, Udit and Wu, Carole-Jean and Azzolini, Alisson G. and Dzhulgakov, Dmytro and Mallevich, Andrey and Cherniavskii, Ilia and Lu, Yinghai and Krishnamoorthi, Raghuraman and Yu, Ansha and Kondratenko, Volodymyr and Pereira, Stephanie and Chen, Xianjie and Chen, Wenlin and Rao, Vijay and Jia, Bill and Xiong, Liang and Smelyanskiy, Misha},
  journal = {arXiv preprint arXiv:1906.00091},
  year = {2019}
}

@inproceedings{covington2016youtube,
  title = {Deep Neural Networks for YouTube Recommendations},
  author = {Covington, Paul and Adams, Jay and Sargin, Emre},
  booktitle = {Proceedings of the 10th ACM Conference on Recommender Systems},
  pages = {191--198},
  year = {2016},
  doi = {10.1145/2959100.2959190}
}

@inproceedings{yi2019sampling,
  title = {Sampling-Bias-Corrected Neural Modeling for Large Corpus Item Recommendations},
  author = {Yi, Xinyang and Yang, Ji and Hong, Lichan and Cheng, Derek Zhiyuan and Heldt, Lukasz and Kumthekar, Aditee Ajit and Zhao, Zhe and Wei, Li and Chi, Ed H.},
  booktitle = {Proceedings of the 13th ACM Conference on Recommender Systems},
  pages = {269--277},
  year = {2019},
  doi = {10.1145/3298689.3346996}
}

@inproceedings{li2021prefix,
  title = {Prefix-Tuning: Optimizing Continuous Prompts for Generation},
  author = {Li, Xiang Lisa and Liang, Percy},
  booktitle = {Proceedings of the 59th Annual Meeting of the Association for Computational Linguistics and the 11th International Joint Conference on Natural Language Processing},
  pages = {4582--4597},
  year = {2021},
  doi = {10.18653/v1/2021.acl-long.353}
}

@inproceedings{lester2021prompt,
  title = {The Power of Scale for Parameter-Efficient Prompt Tuning},
  author = {Lester, Brian and Al-Rfou, Rami and Constant, Noah},
  booktitle = {Proceedings of the 2021 Conference on Empirical Methods in Natural Language Processing},
  pages = {3045--3059},
  year = {2021},
  doi = {10.18653/v1/2021.emnlp-main.243}
}

@inproceedings{li2023blip2,
  title = {BLIP-2: Bootstrapping Language-Image Pre-training with Frozen Image Encoders and Large Language Models},
  author = {Li, Junnan and Li, Dongxu and Savarese, Silvio and Hoi, Steven C. H.},
  booktitle = {Proceedings of the 40th International Conference on Machine Learning},
  pages = {19730--19742},
  year = {2023}
}

@article{wu2023survey,
  title = {A Survey on Large Language Models for Recommendation},
  author = {Wu, Likang and Zheng, Zhi and Qiu, Zhaopeng and Wang, Hao and Gu, Hongchao and Shen, Tingjia and Qin, Chuan and Zhu, Chen and Zhu, Hengshu and Liu, Qi and Xiong, Hui and Chen, Enhong},
  journal = {arXiv preprint arXiv:2305.19860},
  year = {2023}
}

@article{xu2024tapping,
  title = {Tapping the Potential of Large Language Models as Recommender Systems: A Comprehensive Framework and Empirical Analysis},
  author = {Xu, Lanling and Zhang, Junjie and Li, Bingqian and Wang, Jinpeng and Chen, Sheng and Zhao, Wayne Xin and Wen, Ji-Rong},
  journal = {arXiv preprint arXiv:2401.04997},
  year = {2024}
}

@inproceedings{ren2023rlmrec,
  title = {Representation Learning with Large Language Models for Recommendation},
  author = {Ren, Xubin and Wei, Wei and Xia, Lianghao and Su, Lixin and Cheng, Suqi and Wang, Junfeng and Yin, Dawei and Huang, Chao},
  booktitle = {Proceedings of the ACM Web Conference 2024},
  pages = {3464--3475},
  year = {2024},
  doi = {10.1145/3589334.3645458}
}

@inproceedings{wei2024llmrec,
  title = {LLMRec: Large Language Models with Graph Augmentation for Recommendation},
  author = {Wei, Wei and Ren, Xubin and Tang, Jiabin and Wang, Qinyong and Su, Lixin and Cheng, Suqi and Wang, Junfeng and Yin, Dawei and Huang, Chao},
  booktitle = {Proceedings of the 17th ACM International Conference on Web Search and Data Mining},
  pages = {806--815},
  year = {2024},
  doi = {10.1145/3616855.3635853}
}

@article{lyu2023llmrec,
  title = {LLM-Rec: Personalized Recommendation via Prompting Large Language Models},
  author = {Lyu, Hanjia and Jiang, Song and Zeng, Hanqing and Xia, Yinglong and Wang, Qifan and Zhang, Si and Chen, Ren and Leung, Christopher and Tang, Jiajie and Luo, Jiebo},
  journal = {arXiv preprint arXiv:2307.15780},
  year = {2023}
}

@article{li2023generative,
  title = {Large Language Models for Generative Recommendation: A Survey and Visionary Discussions},
  author = {Li, Lei and Zhang, Yongfeng and Liu, Dugang and Chen, Li},
  journal = {arXiv preprint arXiv:2309.01157},
  year = {2023}
}

@inproceedings{rajput2023tiger,
  title = {Recommender Systems with Generative Retrieval},
  author = {Rajput, Shashank and Mehta, Nikhil and Singh, Anima and Keshavan, Raghunandan H. and Vu, Trung and Heldt, Lukasz and Hong, Lichan and Tay, Yi and Tran, Vinh Q. and Samost, Jonah and Kula, Maciej and Chi, Ed H. and Sathiamoorthy, Maheswaran},
  booktitle = {Advances in Neural Information Processing Systems},
  volume = {36},
  pages = {10299--10315},
  year = {2023}
}

@inproceedings{liao2024llara,
  title = {LLaRA: Large Language-Recommendation Assistant},
  author = {Liao, Jiayi and Li, Sihang and Yang, Zhengyi and Wu, Jiancan and Yuan, Yancheng and Wang, Xiang and He, Xiangnan},
  booktitle = {Proceedings of the 47th International ACM SIGIR Conference on Research and Development in Information Retrieval},
  pages = {1785--1795},
  year = {2024},
  doi = {10.1145/3626772.3657690}
}

@inproceedings{lian2018xdeepfm,
  title = {xDeepFM: Combining Explicit and Implicit Feature Interactions for Recommender Systems},
  author = {Lian, Jianxun and Zhou, Xiaohuan and Zhang, Fuzheng and Chen, Zhongxia and Xie, Xing and Sun, Guangzhong},
  booktitle = {Proceedings of the 24th ACM SIGKDD International Conference on Knowledge Discovery and Data Mining},
  pages = {1754--1763},
  year = {2018}
}

@inproceedings{zhou2018din,
  title = {Deep Interest Network for Click-Through Rate Prediction},
  author = {Zhou, Guorui and Song, Chengru and Zhu, Xiaoqiang and Ma, Xiao and Yan, Yanghui and Dai, Xingya and Zhu, Han and Jin, Junqi and Li, Han and Gai, Kun},
  booktitle = {Proceedings of the 24th ACM SIGKDD International Conference on Knowledge Discovery and Data Mining},
  pages = {1059--1068},
  year = {2018}
}

@inproceedings{zhou2019dien,
  title = {Deep Interest Evolution Network for Click-Through Rate Prediction},
  author = {Zhou, Guorui and Mou, Na and Fan, Ying and Pi, Qi and Bian, Weijie and Zhou, Chang and Zhu, Xiaoqiang and Gai, Kun},
  booktitle = {Proceedings of the AAAI Conference on Artificial Intelligence},
  volume = {33},
  pages = {5941--5948},
  year = {2019},
  doi = {10.1609/aaai.v33i01.33015941}
}

@inproceedings{huang2019fibinet,
  title = {FiBiNET: Combining Feature Importance and Bilinear Feature Interaction for Click-Through Rate Prediction},
  author = {Huang, Tongwen and Zhang, Zhiqi and Zhang, Junlin},
  booktitle = {Proceedings of the 13th ACM Conference on Recommender Systems},
  pages = {169--177},
  year = {2019}
}

@inproceedings{song2019autoint,
  title = {AutoInt: Automatic Feature Interaction Learning via Self-Attentive Neural Networks},
  author = {Song, Weiping and Shi, Chence and Xiao, Zhiping and Duan, Zhijian and Xu, Yewen and Zhang, Ming and Tang, Jian},
  booktitle = {Proceedings of the 28th ACM International Conference on Information and Knowledge Management},
  pages = {1161--1170},
  year = {2019}
}

@article{huang2020tabtransformer,
  title = {TabTransformer: Tabular Data Modeling Using Contextual Embeddings},
  author = {Huang, Xin and Khetan, Ashish and Cvitkovic, Milan and Karnin, Zohar},
  journal = {arXiv preprint arXiv:2012.06678},
  year = {2020}
}

@inproceedings{gorishniy2021rtdl,
  title = {Revisiting Deep Learning Models for Tabular Data},
  author = {Gorishniy, Yury and Rubachev, Ivan and Khrulkov, Valentin and Babenko, Artem},
  booktitle = {Advances in Neural Information Processing Systems},
  year = {2021}
}

@inproceedings{alayrac2022flamingo,
  title = {Flamingo: A Visual Language Model for Few-Shot Learning},
  author = {Alayrac, Jean-Baptiste and others},
  booktitle = {Advances in Neural Information Processing Systems},
  year = {2022}
}

@inproceedings{tancik2020fourier,
  title = {Fourier Features Let Networks Learn High Frequency Functions in Low Dimensional Domains},
  author = {Tancik, Matthew and Srinivasan, Pratul P. and Mildenhall, Ben and Fridovich-Keil, Sara and Raghavan, Nithin and Singhal, Utkarsh and Ramamoorthi, Ravi and Barron, Jonathan T. and Ng, Ren},
  booktitle = {Advances in Neural Information Processing Systems},
  volume = {33},
  pages = {7537--7547},
  year = {2020}
}

@inproceedings{mildenhall2020nerf,
  title = {NeRF: Representing Scenes as Neural Radiance Fields for View Synthesis},
  author = {Mildenhall, Ben and Srinivasan, Pratul P. and Tancik, Matthew and Barron, Jonathan T. and Ramamoorthi, Ravi and Ng, Ren},
  booktitle = {European Conference on Computer Vision},
  pages = {405--421},
  year = {2020}
}

@inproceedings{vaswani2017attention,
  title = {Attention Is All You Need},
  author = {Vaswani, Ashish and Shazeer, Noam and Parmar, Niki and Uszkoreit, Jakob and Jones, Llion and Gomez, Aidan N. and Kaiser, Lukasz and Polosukhin, Illia},
  booktitle = {Advances in Neural Information Processing Systems},
  year = {2017}
}

@inproceedings{rajput2023recommender,
  title = {Recommender Systems with Generative Retrieval},
  author = {Rajput, Shashank and Mehta, Nikhil and Singh, Anima and Keshavan, Raghunandan H. and Vu, Trung and Heldt, Lukasz and Hong, Lichan and Tay, Yi and Tran, Vinh Q. and Samost, Jonah and Kula, Maciej and Chi, Ed H. and Sathiamoorthy, Maheswaran},
  booktitle = {Advances in Neural Information Processing Systems},
  volume = {36},
  pages = {10299--10315},
  year = {2023}
}

@inproceedings{he2016ups,
  title = {Ups and Downs: Modeling the Visual Evolution of Fashion Trends with One-Class Collaborative Filtering},
  author = {He, Ruining and McAuley, Julian},
  booktitle = {Proceedings of the 25th International Conference on World Wide Web},
  series = {WWW '16},
  pages = {507--517},
  year = {2016},
  publisher = {International World Wide Web Conferences Steering Committee},
  doi = {10.1145/2872427.2883037}
}

@article{yang2025qwen3,
  title = {Qwen3 Technical Report},
  author = {Yang, An and Li, Anfeng and Yang, Baosong and Zhang, Beichen and Hui, Binyuan and Zheng, Bo and Yu, Bowen and Gao, Chang and Huang, Chengen and Lv, Chenxu and Zheng, Chujie and Liu, Dayiheng and Zhou, Fan and Huang, Fei and Hu, Feng and Ge, Hao and Wei, Haoran and Lin, Huan and Tang, Jialong and Yang, Jian and Tu, Jianhong and Zhang, Jianwei and Yang, Jianxin and Yang, Jiaxi and Zhou, Jing and Zhou, Jingren and Lin, Junyang and Dang, Kai and Bao, Keqin and Yang, Kexin and Yu, Le and Deng, Lianghao and Li, Mei and Xue, Mingfeng and Li, Mingze and Zhang, Pei and Wang, Peng and Zhu, Qin and Men, Rui and Gao, Ruize and Liu, Shixuan and Luo, Shuang and Li, Tianhao and Tang, Tianyi and Yin, Wenbiao and Ren, Xingzhang and Wang, Xinyu and Zhang, Xinyu and Ren, Xuancheng and Fan, Yang and Su, Yang and Zhang, Yichang and Zhang, Yinger and Wan, Yu and Liu, Yuqiong and Wang, Zekun and Cui, Zeyu and Zhang, Zhenru and Zhou, Zhipeng and Qiu, Zihan},
  journal = {arXiv preprint arXiv:2505.09388},
  year = {2025},
  doi = {10.48550/arXiv.2505.09388}
}

@inproceedings{hidasi2016gru4rec,
  title     = {Session-based Recommendations with Recurrent Neural Networks},
  author    = {Hidasi, Bal{\'a}zs and Karatzoglou, Alexandros and Baltrunas, Linas and Tikk, Domonkos},
  booktitle = {International Conference on Learning Representations},
  year      = {2016},
  url       = {https://arxiv.org/abs/1511.06939}
}

@inproceedings{tang2018caser,
  title     = {Personalized Top-N Sequential Recommendation via Convolutional Sequence Embedding},
  author    = {Tang, Jiaxi and Wang, Ke},
  booktitle = {Proceedings of the Eleventh ACM International Conference on Web Search and Data Mining},
  pages     = {565--573},
  year      = {2018},
  publisher = {ACM},
  doi       = {10.1145/3159652.3159656}
}

@inproceedings{ma2019hgn,
  title     = {Hierarchical Gating Networks for Sequential Recommendation},
  author    = {Ma, Chen and Kang, Peng and Liu, Xue},
  booktitle = {Proceedings of the 25th ACM SIGKDD International Conference on Knowledge Discovery and Data Mining},
  pages     = {825--833},
  year      = {2019},
  publisher = {ACM},
  doi       = {10.1145/3292500.3330984}
}

@inproceedings{kang2018sasrec,
  title     = {Self-Attentive Sequential Recommendation},
  author    = {Kang, Wang-Cheng and McAuley, Julian},
  booktitle = {2018 IEEE International Conference on Data Mining},
  pages     = {197--206},
  year      = {2018},
  publisher = {IEEE},
  doi       = {10.1109/ICDM.2018.00035}
}

@inproceedings{sun2019bert4rec,
  title     = {{BERT4Rec}: Sequential Recommendation with Bidirectional Encoder Representations from Transformer},
  author    = {Sun, Fei and Liu, Jun and Wu, Jian and Pei, Changhua and Lin, Xiao and Ou, Wenwu and Jiang, Peng},
  booktitle = {Proceedings of the 28th ACM International Conference on Information and Knowledge Management},
  pages     = {1441--1450},
  year      = {2019},
  publisher = {ACM},
  doi       = {10.1145/3357384.3357895}
}

@inproceedings{zhang2019fdsa,
  title     = {Feature-level Deeper Self-Attention Network for Sequential Recommendation},
  author    = {Zhang, Tingting and Zhao, Pengpeng and Liu, Yanchi and Sheng, Victor S. and Xu, Jiajie and Wang, Deqing and Liu, Guanfeng and Zhou, Xiaofang},
  booktitle = {Proceedings of the 28th International Joint Conference on Artificial Intelligence},
  pages     = {4320--4326},
  year      = {2019},
  doi       = {10.24963/ijcai.2019/600}
}

@inproceedings{zhou2020s3rec,
  title     = {{S3-Rec}: Self-Supervised Learning for Sequential Recommendation with Mutual Information Maximization},
  author    = {Zhou, Kun and Wang, Hui and Zhao, Wayne Xin and Zhu, Yutao and Wang, Sirui and Zhang, Fuzheng and Wang, Zhongyuan and Wen, Ji-Rong},
  booktitle = {Proceedings of the 29th ACM International Conference on Information and Knowledge Management},
  pages     = {1893--1902},
  year      = {2020},
  publisher = {ACM},
  doi       = {10.1145/3340531.3411954}
}

\clearpage
\beginappendix
\section{Data Preprocessing}
\label{sec:appendix-preprocessing}

We convert the Amazon review data into natural-language prompts for both user-side and item-side inputs. In implementation, we construct two prompt granularities. The first, a concise \emph{summarized prompt}, is used as the direct textual input to the final Qwen3-0.6B tower. The second, a \emph{rich prompt}, preserves more detailed textual and numerical information for constructing the auxiliary embedding and numerical features.

\subsection{Summarized Prompt}

The summarized prompt provides compact user and item descriptions for the final retrieval tower.

\paragraph{User prompt.}
For each user, the summarized prompt contains:
\begin{itemize}
    \item \textbf{Recent items.} The five most recent items with which the user interacted, each formatted as ``[Brand] Title (LeafCategory, L2Category) Price Rating''.
    \item \textbf{Taste summary.} For users with at least three historical items, we summarize the top three second-level (L2) categories with their percentage shares, retaining only categories whose share is at least 15\%. We also include the top two brands and the observed price range.
\end{itemize}

\paragraph{Item prompt.}
For each item, the summarized prompt contains:
\begin{itemize}
    \item \textbf{Core information.} ASIN, title, leaf category, second-level (L2) category, brand, price, average rating, and number of ratings.
    \item \textbf{Description snippet.} The first 100 characters of the product description, truncated at a word boundary.
    \item \textbf{Also-bought signal.} Up to three co-purchased items, formatted as ``Also bought: [Brand] Title (LeafCategory) | [Brand] Title (LeafCategory) | ...''.
\end{itemize}

\subsection{Rich Prompt}

The rich prompt is used to construct the embedding and numerical feature streams. Compared with the summarized prompt, it preserves more detailed behavioral signals, metadata, and review information.

\paragraph{User prompt.}
The rich user prompt contains two components:
\begin{itemize}
    \item \textbf{Taste profile.} The taste profile is aggregated over all historical items visible during training. It includes the top four second-level categories and their percentage shares, retaining categories whose share is at least 5\%; the top three most frequent brands; the minimum and maximum prices across historical items; and the average rating of items purchased by the user.
    \item \textbf{Recent interactions.} The five most recent historical items are listed. Each item is formatted as ``[Brand] Title (leaf category) | price | rating''.
\end{itemize}

\paragraph{Item prompt.}
The rich item prompt contains the following fields for the target item:
\begin{itemize}
    \item \textbf{Header.} ASIN and title, with the title truncated to 100 characters.
    \item \textbf{Attributes.} Brand, full category path, price rounded to two decimal places, average rating, and number of reviews.
    \item \textbf{Description.} Product description truncated to at most 400 characters.
    \item \textbf{Buyer comments.} Up to five review summaries, sorted by review rating in descending order and enclosed in quotation marks.
    \item \textbf{Also-bought items.} Up to three titles from the \texttt{bought\_together} and \texttt{also\_bought} metadata fields, resolved to item titles when available.
    \item \textbf{Also-viewed items.} Up to three titles from the \texttt{also\_viewed} metadata field, excluding items already included in the also-bought list.
\end{itemize}

\section{Metrics}
\label{sec:appendix-metrics}

\paragraph{Recall@K.}
Let $\mathcal{U}$ denote the set of users, $i_u^{\ast}$ the ground-truth next item for user $u \in \mathcal{U}$, and $\mathcal{R}_u^{K}$ the set of top-$K$ items ranked by the model. Recall@K is defined as
\begin{equation}
\mathrm{Recall@}K
= \mathbb{E}_{u\in\mathcal{U}}\left[
\mathbb{I}\!\left( i_u^{\ast} \in \mathcal{R}_u^{K} \right)
\right],
\end{equation}
where $\mathbb{I}(\cdot)$ is the indicator function.

\paragraph{NDCG@K.}
For user $u$, the Discounted Cumulative Gain at $K$ (DCG@K) is computed as
\begin{equation}
\mathrm{DCG@}K
= \sum_{j=1}^{K}
\frac{2^{\mathrm{rel}_{u,j}} - 1}{\log_2(j+1)},
\end{equation}
where $\mathrm{rel}_{u,j} \in \{0,1\}$ indicates whether the item ranked at position $j$ matches the ground-truth next item $i_u^{\ast}$.

Under the leave-one-out evaluation protocol, each user has a single ground-truth item. The ideal ranking therefore places this item at the first position, giving $\mathrm{IDCG@}K=1$. NDCG@K is defined as
\begin{equation}
\mathrm{NDCG@}K
= \mathbb{E}_{u\in\mathcal{U}}\left[
\frac{\mathrm{DCG@}K}{\mathrm{IDCG@}K}
\right].
\end{equation}

\section{Hyperparameter Study}
\label{sec:appendix-hyperparameter}

We conduct a hyperparameter study for the token-fusion module in Eq.~\eqref{eq:token-fusion}. Specifically, we compare bidirectional attention and two one-way variants, numerical $\rightarrow$ embedding and embedding $\rightarrow$ numerical, using either one or two fusion layers. Tables~\ref{tab:hyperparam-beauty}--\ref{tab:hyperparam-toys} report the results on Beauty, Sports and Outdoors, and Toys and Games, respectively. The optimal configuration is dataset-dependent: bidirectional attention with two layers performs best on Beauty, embedding $\rightarrow$ numerical attention with one layer performs best on Sports and Outdoors, and embedding $\rightarrow$ numerical attention with two layers performs best on Toys and Games.

\begin{table}[!t]
\centering
\caption{Configuration search on the Beauty dataset.}
\label{tab:hyperparam-beauty}
\begin{tabular}{lccccc}
\toprule
\textbf{Attention Direction} & \textbf{\# of Layers} & \textbf{R@5} & \textbf{R@10} & \textbf{N@5} & \textbf{N@10} \\
\midrule
Bidirectional & 1 & 0.0406 & 0.0724 & 0.0239 & 0.0341 \\
Bidirectional & 2 & \textbf{0.0459} & \textbf{0.0766} & \textbf{0.0278} & \textbf{0.0377} \\
Numerical $\rightarrow$ Embedding & 1 & 0.0370 & 0.0653 & 0.0218 & 0.0309 \\
Numerical $\rightarrow$ Embedding & 2 & 0.0443 & 0.0758 & 0.0267 & 0.0368 \\
Embedding $\rightarrow$ Numerical & 1 & 0.0404 & 0.0717 & 0.0236 & 0.0336 \\
Embedding $\rightarrow$ Numerical & 2 & 0.0452 & 0.0722 & 0.0273 & 0.0359 \\
\bottomrule
\end{tabular}
\end{table}

\begin{table}[!t]
\centering
\caption{Configuration search on the Sports and Outdoors dataset.}
\label{tab:hyperparam-sports}
\begin{tabular}{lccccc}
\toprule
\textbf{Attention Direction} & \textbf{\# of Layers} & \textbf{R@5} & \textbf{R@10} & \textbf{N@5} & \textbf{N@10} \\
\midrule
Bidirectional & 1 & 0.0223 & 0.0371 & 0.0134 & 0.0182 \\
Bidirectional & 2 & 0.0197 & 0.0340 & 0.0116 & 0.0162 \\
Numerical $\rightarrow$ Embedding & 1 & 0.0247 & 0.0407 & 0.0147 & 0.0198 \\
Numerical $\rightarrow$ Embedding & 2 & 0.0217 & 0.0360 & 0.0137 & 0.0182 \\
Embedding $\rightarrow$ Numerical & 1 & \textbf{0.0253} & \textbf{0.0410} & \textbf{0.0154} & \textbf{0.0205} \\
Embedding $\rightarrow$ Numerical & 2 & 0.0217 & 0.0371 & 0.0134 & 0.0183 \\
\bottomrule
\end{tabular}
\end{table}

\begin{table}[!t]
\centering
\caption{Configuration search on the Toys and Games dataset.}
\label{tab:hyperparam-toys}
\begin{tabular}{lccccc}
\toprule
\textbf{Attention Direction} & \textbf{\# of Layers} & \textbf{R@5} & \textbf{R@10} & \textbf{N@5} & \textbf{N@10} \\
\midrule
Bidirectional & 1 & 0.0651 & 0.1024 & 0.0397 & 0.0517 \\
Bidirectional & 2 & 0.0627 & 0.0995 & 0.0363 & 0.0482 \\
Numerical $\rightarrow$ Embedding & 1 & 0.0585 & 0.0974 & 0.0341 & 0.0466 \\
Numerical $\rightarrow$ Embedding & 2 & 0.0642 & 0.1023 & 0.0384 & 0.0507 \\
Embedding $\rightarrow$ Numerical & 1 & 0.0622 & 0.0997 & 0.0371 & 0.0492 \\
Embedding $\rightarrow$ Numerical & 2 & \textbf{0.0695} & \textbf{0.1064} & \textbf{0.0417} & \textbf{0.0536} \\
\bottomrule
\end{tabular}
\end{table}

\section{Additional Ablation Study}
\label{sec:appendix-additional-ablation}

The main ablations in Section~\ref{sec:exp} show that interaction-based fusion is preferable to direct concatenation of heterogeneous soft tokens. We further examine three design alternatives for the components introduced in Section~\ref{sec:method}: using Q-Former adapters to construct fixed-length soft tokens, using Fourier features to encode continuous prices, and modeling the two prompt granularities through soft-token interaction rather than naive text concatenation. Using the selected interaction-based fusion model as the reference, A1 replaces the Q-Former adapters with MLP projections, A2 replaces Fourier price encoding with a tokenizer-based numerical representation, and A3 directly concatenates the summarized and rich prompts into a single LLM input. All other training and evaluation settings are kept unchanged.

\begin{table}[!t]
\centering
\caption{Additional design ablations on the Beauty dataset. $\Delta$R@5 is the relative change from interaction-based fusion.}
\label{tab:ablation-beauty}
\begin{tabular}{lccccc}
\toprule
Variant & R@5 & R@10 & N@5 & N@10 & $\Delta$R@5 \\
\midrule
Interaction-based fusion & \textbf{0.0459} & \textbf{0.0766} & \textbf{0.0278} & \textbf{0.0377} & --- \\
A1 Q-Former$\to$MLP & 0.0420 & 0.0734 & 0.0240 & 0.0341 & $-8.5\%$ \\
A2 Fourier$\to$tokenizer & 0.0416 & 0.0702 & 0.0248 & 0.0340 & $-9.3\%$ \\
A3 Prompt concatenation & 0.0419 & 0.0715 & 0.0246 & 0.0341 & $-8.7\%$ \\
\bottomrule
\end{tabular}
\end{table}

\begin{table}[!t]
\centering
\caption{Additional design ablations on the Sports and Outdoors dataset. $\Delta$R@5 is the relative change from interaction-based fusion.}
\label{tab:ablation-sports}
\begin{tabular}{lccccc}
\toprule
Variant & R@5 & R@10 & N@5 & N@10 & $\Delta$R@5 \\
\midrule
Interaction-based fusion & \textbf{0.0253} & \textbf{0.0410} & \textbf{0.0154} & \textbf{0.0205} & --- \\
A1 Q-Former$\to$MLP & 0.0101 & 0.0177 & 0.0060 & 0.0085 & $-60.3\%$ \\
A2 Fourier$\to$tokenizer & 0.0208 & 0.0379 & 0.0120 & 0.0175 & $-17.8\%$ \\
A3 Prompt concatenation & 0.0212 & 0.0383 & 0.0120 & 0.0175 & $-16.2\%$ \\
\bottomrule
\end{tabular}
\end{table}

\begin{table}[!t]
\centering
\caption{Additional design ablations on the Toys and Games dataset. $\Delta$R@5 is the relative change from interaction-based fusion.}
\label{tab:ablation-toys}
\begin{tabular}{lccccc}
\toprule
Variant & R@5 & R@10 & N@5 & N@10 & $\Delta$R@5 \\
\midrule
Interaction-based fusion & \textbf{0.0695} & \textbf{0.1064} & \textbf{0.0417} & \textbf{0.0536} & --- \\
A1 Q-Former$\to$MLP & 0.0584 & 0.0968 & 0.0341 & 0.0464 & $-15.9\%$ \\
A2 Fourier$\to$tokenizer & 0.0623 & 0.0983 & 0.0365 & 0.0481 & $-10.3\%$ \\
A3 Prompt concatenation & 0.0625 & 0.0992 & 0.0383 & 0.0502 & $-10.1\%$ \\
\bottomrule
\end{tabular}
\end{table}

Across all three datasets, every ablated variant reduces retrieval quality relative to the interaction-based fusion model. Replacing Q-Former adapters with MLP projections lowers R@5 by 8.5\%, 60.3\%, and 15.9\% on Beauty, Sports and Outdoors, and Toys and Games, respectively, indicating that query-based soft-token compression is important for preserving useful information from heterogeneous feature streams. Replacing Fourier price encoding with a tokenizer-based numerical representation also consistently degrades R@5, with relative drops of 9.3\%, 17.8\%, and 10.3\%. The naive prompt-concatenation baseline reduces R@5 by 8.7\%, 16.2\%, and 10.1\%, showing that simply concatenating textual prompts is not sufficient to replace explicit soft-token construction and interaction. These results provide empirical support for adopting Q-Former soft-token adapters, Fourier numerical encodings, and interaction-based soft-token fusion in the proposed method.

\end{document}